\documentclass[journal,letterpaper]{IEEEtran}
\usepackage{cite}
\usepackage{graphicx}
\begin{document}
\title{Dynamic hysteresis in Finemet thin films}
\author{L.~Santi, R.L. Sommer, A. Magni, G. Durin, F. Colaiori, and S. Zapperi%
\thanks{L. Santi and R.L. Sommer are with the Univ. Federal de Santa Maria, Dep. de Fisica UFSM,
97105-900, Santa Maria, RS, Brasil. Email: santi@ien.it, sommer@ccne.ufsm.br}%
\thanks{ A. Magni and  G. Durin are with the IEN Galileo Ferraris, str. delle Cacce 91, 10135 Torino,
Italy. Email: magni@ien.it, durin@ien.it}%
\thanks{F. Colaiori and S. Zapperi are with the INFM Unit\`{a} di Roma 1,
Dip. di Fisica, Universit\`{a} "La Sapienza",  P.le A. Moro 2, 00185 Roma, Italy.
Email: fran@pil.phys.uniroma1.it, zapperi@pil.phys.uniroma1.it}}

\maketitle

\begin{abstract}
We performed a series of dynamic hysteresis measurements on three series of Finemet
films with composition Fe$_{73.5}$Cu$_1$Nb$_3$Si$_13.5$B$_9$,
using both the longitudinal magneto-optical Kerr effect (MOKE) and the inductive
fluxometric method. The MOKE dynamic hysteresis loops show a more marked variability
with the frequency than the inductive ones, while both measurements show a similar
dependence on the square root of frequency. We analyze these results in the frame of a
simple domain wall depinning model, which accounts for the general behavior of the
data.
\end{abstract}


\section{Introduction}
The physics of thin and ultrathin magnetic films has been extensively studied in the
recent past, because of its great importance in several applications, ranging from
multilayers to high frequency devices. For this reason, many recent papers have been
devoted to measure the magnetization reversal dynamics in two dimensional structures,
revealing the existence of universal features and scale-invariant properties of the
hysteresis loops \cite{RUI-02,CHO-99,JIA-95,HE-93}. Despite these efforts, a general
description of these features is still an open problem, as most experimental results
are still to be interpreted in the framework of the existing models
\cite{ZHO-02,LYU-99,CHA-99,LEE-99}. In particular, the dynamic hysteresis loop area $A$
is often assumed to scale as $A \propto H_0^\alpha \omega^\beta T^{-\gamma}$, where
$H_0$ is the amplitude of sinusoidal external field of frequency $\omega$, $T$ is the
temperature, and $\alpha$, $\beta$, $\gamma$ are three scaling exponents. As a matter
of fact, experimental evaluation of the exponents $\alpha$ and $\beta$ in the low
dynamic regime spans a quite large range from 0 to 0.8, with a general higher value for
thinner films \cite{LEE-99,MOO-01} (see also Tab. I of \cite{SUE-97}). The proposed
theoretical models roughly span the same range, so that a clear identification of the
fundamental properties of magnetization dynamics seems far to be reached. In order to
investigate this complicated problem offering a new perspective, we present a series of
dynamic hysteresis measurements on Finemet thin films by using the magneto-optical Kerr
effect (MOKE), as employed in all the studies presented in the literature, and the
fluxmetric inductive method, using a pick-up coil wound around the sample. This enables
us to investigate the hysteresis properties not only considering the magnetization
changes of the surface within the laser spot area, but also those of the total volume
of the sample; this is particularly important in order to check the dependence of the
loop area on the film thickness, and to understand the true nature of the magnetization
dynamics. Quite unexpectedly, MOKE hysteresis loops show a remarkable variability with
the frequency, about one order of magnitude higher than the inductive ones; on the
other hand, both methods give a similar dependence on the frequency. We try to
interpret these results with a simple domain wall depinning model which can be solved
analytically, giving reason of the general behavior of the data.

\section{Materials and measurement methods}
Films having nominal composition Fe$_{73.5}$Cu$_1$Nb$_3$Si$_13.5$B$_9$ have been
deposited on glass substrates by rf magnetron sputtering under a 5 mTorr Ar atmosphere
at room temperature. The sample thickness, ranging from about 21 nm to 5 $\mu$m, is
measured by angle X-ray diffractometry, which also confirms the amorphous state of the
samples. The hysteresis loops of the in-plane magnetization are measured in the
as-prepared materials as a function of the applied field (up to a few kA/m) at 100 Hz,
and as a function of the frequency. The longitudinal MOKE measurements are performed
with an optical bench equipped with an He-Ne laser light source, covering a sample
surface of about 1 mm$^2$, and a photodiode having a frequency cutoff above 150kHz.
Samples are cut to a maximum size of about 2 x 2 mm and inserted in a Helmholtz coil
setup giving a maximum field of about 15 kA/m. In this configuration, we could perform
measurements up to 300 Hz.

The fluxometric measurements of hysteresis loops are performed on larger samples,
usually cut within an homogeneous region of 3 by 1.5 cm. For these samples, we prepared
a 10 cm long solenoid with 720 coils and a N/L value of 13200. The sinusoidal applied
field is measured detecting the voltage over a calibrated 1 $\Omega$ resistance. The
induced flux is detected with two sets of 50 coils wound around the sample and covering
an area of about 1 cm; the two sets are wired in order to cancel out the air flux and
directly detect the film magnetization changes. As the film cross section is much
smaller than the area of the coils, a perfect cancellation of the air flux is often
difficult, requiring a continuous adjustment of area of one of the sets. This problem
further increases at high frequencies, due to the different coupling between the wires,
making it hard to perform the measurement. We thus get the full cancellation by
numerically subtracting a sinusoidal wave with proper amplitude and phase from the
induced signal. This procedure has a certain degree of arbitrariness, as small changes
of the sinusoidal amplitude and phase give slightly different loop shapes. In all
cases, the loop area and the coercive field are not substantially affected.

\begin{figure}
\centering \includegraphics[width=3in]{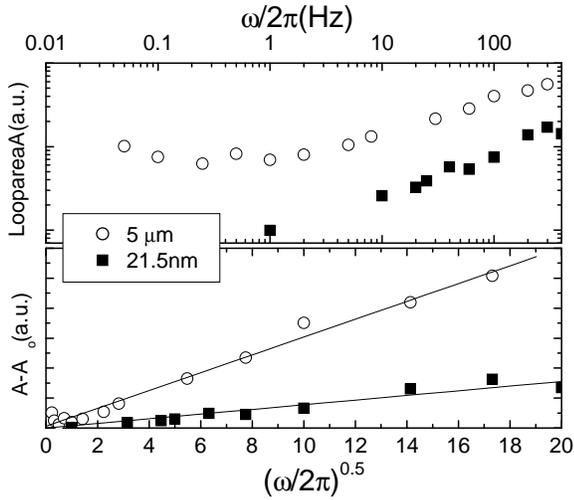}

\caption{MOKE measurements of the hysteresis loop area $A$ (top) for the thickest (5
$\mu$m - open dots) and the thinner  (21 nm - squares) Finemet sample. Data for the 5
$\mu$m  would suggest the presence of a dynamic transition around a few Hz. A plot of
the area after subtraction of the static value $A_0$ (bottom) shows instead a linear
dependence with the square root of frequency ($\beta = 0.5$ in~(\ref{eq:A})).}
\label{fig:moke}
\end{figure}

\section{Experimental results}
The first fundamental result of both kind of measurements is the existence of a well
defined static hysteresis, as usually found in magnetic materials (see for instance
\cite{Bertotti}). Therefore, the loop area $A$ is better described by
\begin{equation}\label{eq:A}
     A \sim A_0 + H_0^\alpha \omega^\beta T^{-\gamma}
\end{equation}
where the static loop area $A_0$ is estimated using data at the lowest frequencies.
Clearly, the choice of (\ref{eq:A}) completely changes the experimental estimation of
critical exponents. We believe that this simple observation could explain the so called
dynamic transition, a sharp change in the value of the exponent $\beta$ at intermediate
frequencies. Due to the usual large value of $A_0$ with respect to variation of the
data with the frequency, log-log plots can mimic a dynamic transition when data
actually follow a simple power law as assumed in (\ref{eq:A}). Our results and the
theoretical analysis should help to clarify this important point.

The MOKE hysteresis loop area for different sample thickness show large variations with
the frequency, as shown in Fig.~\ref{fig:moke} for the thickest (5 $\mu$m) and thinnest
(21 nm) samples. Visual inspection of this Figure would suggest a dynamic transition
around a few Hz, at least, for the 5 $\mu$m sample. As a matter of fact, the data show
a much simpler behavior: in fact, the loop area $A$ follows a simple law of the type $A
\simeq A_0 + k \omega^{0.5}$, where $A_0$ is estimated using low frequency data.

\begin{figure}
\centering
\includegraphics[width=2.8in]{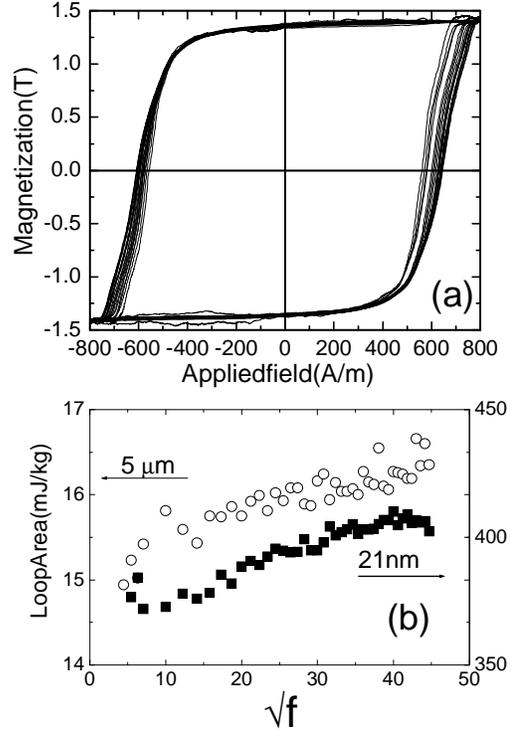}
\caption{(a) Hysteresis loops of a
21 nm Fe$_{73.5}$Cu$_1$Nb$_3$Si$_{13.5}$B$_9$ Finemet film up to 2 kHz, with an
external field amplitude of 800 A/m. (b) Frequency dependence of loop area of the same
sample and of a 5 $\mu$m film: note the approximately linear dependence as a function
of the square root of the frequency.} \label{fig:Loops}
\end{figure}

Surprisingly, hysteresis loops obtained with the fluxometric setup do not show the same
frequency variability. As reported in Figs.~\ref{fig:Loops} for the 21 nm film, the
loop area changes only about 10\% with respect to the static value $A_0$. On the other
hand, when plotted as a function of the square root of frequency, these data follow
reasonably well a linear behavior. This suggests a common mechanism responsible for the
magnetization change, which we try to interpret considering the domain wall depinning
models, as discussed below.

Inductive measurements of the hysteresis loops at 100 Hz as a function of the applied
field amplitude do not show any scaling behavior as given by (\ref{eq:A}) or similar.
Simple squared loops as the ones shown in Fig.~\ref{fig:Loops} are observed only for
the thinner films, while more complicated shapes appear for thicker samples, showing a
clear evidence of multi-domains magnetization processes. We postpone the discussion of
these rich but complex features to a further longer publication.

\section{Model}

\begin{figure}
\centering \includegraphics[width=2.8in]{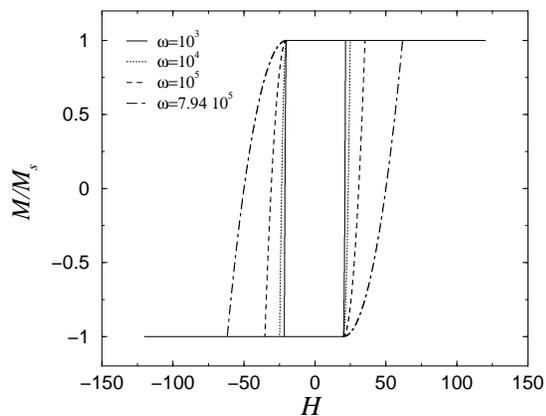} \caption{Dynamic hysteresis
loops calculated using the simple domain wall model of (\ref{eq:v}). Hysteresis
branches are described by (\ref{eq:loopsmallomega}).} \label{fig:sim}
\end{figure}

To understand the frequency dependence of our experimental data, we employ the domain
wall depinning model described in Ref.~\cite{RUI-02}. Under the assumption that
hysteresis is mainly due to domain wall motion, we consider a phenomenological law for
the wall velocity given by
\begin{equation}
v(H) = \mu (|H|-H_p)~ \theta(|H|-H_p), \label{eq:v}
\end{equation}
where $\mu$ the wall mobility, $H_p$ is the depinning field, $\theta$ is the step
function, and the applied field is $H=H_0 sin(\omega t)$. Following ~\cite{LYU-99}, we
solve ~(\ref{eq:v}) and compute the coercive field and the loop area, which at low
frequency are given by
\begin{equation}\label{eq:Hc}
  H_c(\omega) \simeq H_p +(2 L\omega \sqrt{H_0^2-H_p^2}/\mu)^{1/2},
\end{equation}
where $L$ is the sample size, in the case of a single domain wall, or the typical
distance between domain walls in a more general case. The lower branch of the
hysteretic loop is then given by
\begin{equation}
M(h)=
\left\{
    \begin{array}{ll}
    -M_s & H<H_p \\
    M_s \left(
    \frac{\mu}{L\omega}\frac{(H-H_p)^2}{\sqrt{H_0^2-H_p^2}}-1
    \right) & H\geq H_p
    \end{array}
\right. \label{eq:loopsmallomega}
\end{equation}
where the second equation is valid as long as $M<M_s$. Fig.~\ref{fig:sim} displays the
loop shapes for different values of frequency $\omega$. The loop area is thus easily
computed and, for low frequencies, is given by
\begin{equation}
A\simeq A_0+ M_s\frac{8}{3}\sqrt{\frac{2\omega L}{\mu}\sqrt{H_0^2-H_p^2}} \,,
\label{eq:areasmallomega}
\end{equation}
where $A_0=4M_s H_p$ is the area of the loop in the quasistatic limit.


%

\section{Discussion and conclusion}

The simple model shown above gives a simple $\sqrt{\omega}$ dependence of the loop area
(\ref{eq:areasmallomega}), which we believe can account for some experimental data
found in the literature and interpreted assuming the presence of a dynamic transition.
The value $\beta=0.5$ is a strict consequence of the linear form of (\ref{eq:v}): in a
more general case, shown in detail in \cite{LYU-99}, one gets $\beta\geq 0.5$. It is
worth noting that the model above describes the magnetization dynamics of a single
domain wall having a well defined depinning field $H_p$. The model does not include the
effects of any random disorder, the multiplicity of the domains, or the field-dependent
nucleation of domain walls. As a consequence, it cannot describe minor hysteresis loops
or other features, such as more complicated shapes of the loops. Despite these
limitations, we believe it can account for the general behavior of the dynamic
hysteresis, and could be successfully applied to describe thin films showing simple
magnetization dynamics.

Using the results of the model, we found that our experimental data obtained with
complementary measurement techniques are the consequence of the same magnetization
dynamics. That means that the MOKE measurements on the surface are related to those of
the volume, given by the inductive method. While this is expected in thin films, it is
not necessarily valid for the thickest samples, for instance in the case of a few
microns. It is not clear anyway what can account for the large differences in the
change of the loop area. It is worth noting that the MOKE measurement uses a laser spot
area close to the sample dimension so, in principle, it should detect the magnetization
changes of the entire sample, as long as we consider the thinnest films. Therefore, we
should observe similar variations using the two methods. On the other hand, one can
suggest that samples with the same thickness but different later dimensions could not
have the same magnetization dynamics, because of the different role of the
demagnetizing fields, or of the number of domain walls. This possibility has been ruled
out by performing fluxometric measurements on the same sample used in MOKE, having
thickness 5 $\mu$m. In this case the fluxometric measurements show the same behavior
described in Fig.~\ref{fig:Loops}. Unfortunately, measurements at lower thickness are
not possible due to the very low intensity of the induced signal, given the reduce
cross section available. However, we feel confident that this result proves that the
sample lateral dimensions are not relevant, also considering that the data reported in
literature refer to dependences only on the sample thickness and not, more generally,
to the other two dimensions.

In summary, we have shown in this paper that the dynamic hysteresis of
Fe$_{73.5}$Cu$_1$Nb$_3$Si$_13.5$B$_9$ films exhibit different behavior as a function of
sample thickness and magnetizing field frequency, and that the domain wall depinning
model accounts for the general behavior of the data



\end{document}